\documentstyle[preprint,aps,epsfig]{revtex}
\draft
\begin{document}
\bibliographystyle{prsty}
 
\preprint{RUB-TPII-11/95}
\title{Strange vector
form factors of the nucleon in the SU(3) chiral quark-soliton model
with the proper kaonic cloud}

\author{Hyun-Chul Kim
\footnote{E-mail address: kim@hadron.tp2.ruhr-uni-bochum.de},
Teruaki Watabe 
\footnote{E-mail address: watabe@hadron.tp2.ruhr-uni-bochum.de},
and Klaus Goeke
\footnote{E-mail address: goeke@hadron.tp2.ruhr-uni-bochum.de}}

\address{
Institut f\"ur Theoretische  Physik  II, \\  Postfach 102148,
Ruhr-Universit\" at Bochum, \\
 D--44780 Bochum, Germany  \\
       }
\date{March 1996}
\maketitle
\begin{abstract}
The strange vector form factors are evaluated 
in the range between $Q^2=0$ and $Q^2=1\ \mbox{GeV}^2$ in the framework
of the SU(3) chiral quark-soliton model (or semi-bosonized SU(3) 
Nambu-Jona-Lasinio model).   The rotational $1/N_c$ 
and $m_s$ corrections are taken into account up to linear order.  
Taking care of a proper Yukawa-tail of the kaonic cloud,
we get $\langle r^{2}\rangle^{\rm Sachs}_{s}=-0.095\;
\mbox{fm}^2$ and $\mu_s = -0.68\;\mu_N$.
The results are compared with several different models.  
\end{abstract}

\vfill\eject
\section{Introduction}
The strangeness content of the nucleon has been under a great deal 
of discussions for well over a decade.
A few years ago, the European Muon Collaboration (EMC)~\cite{ashman} 
measured the spin structure function of the
proton in deep inelastic muon scattering and
showed that there is an indication of
a sizable strange quark contribution. 
This remarkable result has been confirmed by following experiments
of the Spin Muon Collaboration (SMC)~\cite{SMC1,SMC2},
E142 and E143 collaborations~\cite{E142,E143}.

Another experiment conducted at Brookhaven~\cite{ahrens} 
(BNL experiment 734)
measuring the low-energy elastic neutrino-proton scattering
came to the more or less same conclusion.  
 Kaplan and Manohar~\cite{km} 
showed how elastic $\nu p$ and $ep$ scatterings can be used 
to extract not only the $G_{1}$ form factors of the $U(1)_A$ current
but also the $F_2$ form factors of the baryon number current
and furthermore how the strange quark matrix elements
$\langle p | \bar{s} \gamma_\mu \gamma_5s|p \rangle $ and 
$\langle p | \bar{s} \gamma_\mu s | p \rangle $ can be evaluated.
 Following these suggestions, 
Garvey {\em et al.}~\cite{garvey1} reanalyzed the above-mentioned
$\nu p$ elastic scattering experiment and determined proton
strange form factors in particular at $Q^2=0$, pointing
out the shortcomings of the analysis done by Ref.~\cite{ahrens} .
The best fit of Ref.~\cite{garvey1} with the smallest $\chi^2$ tells
$F^{s}_{1}=0.53\pm 0.70$ and $F^{s}_{2}=-0.40\pm 0.72$.
By comparing the different $Q^2$ dependence of 
$d\sigma/dQ^2(\nu p)$ to $d\sigma/dQ^2(\bar{\nu} p)$,
Garvey {\em et al}. favor $F^{s}_{1} (Q^2) > 0$ and
$F^{s}_{2} (Q^2) < 0$.              
However, these form factors are 
experimentally unknown to date and have no stringent 
and concrete constraints on their $Q^2$--dependence yet.  
There are various proposals and experiments in progress
(see Refs.~\cite{musolfetal,bm} for details).
All these considerations lead to the conclusion that,
in contrast to the naive quark model, it is of great importance
to consider strange quarks in the nucleon seriously.

There have been several theoretical efforts to describe the strange 
form factors of the nucleon.  The first attempt
was performed by Jaffe~\cite{jaffe}.  Jaffe took advantage of
Ref.~\cite{hoehler}, {\em i.e.}
the pole fit analysis based on dispersion theory, and
estimated the mean-square strange radius and magnetic moment
of the nucleon:$\langle r^2\rangle^{\rm Dirac}_{s} 
= 0.16\pm 0.06 \ \mbox{fm}^2$, $\mu_s = -0.31\pm 0.09 \ \mu_N$.
 More recently, 
Hammer {\em et al.}~\cite{Hammeretal} updated Jaffe's pole-fit 
analysis of the strange vector form factors,
relying upon a new dispersion theoretic analysis of the nucleon 
electromagnetic form factors.  In fact, Hammer {\em et al.} improved
Jaffe's prediction, giving $\mu_s = -0.24\pm 0.03 \ 
\mu_N$ and $\langle r^2\rangle^{\rm Dirac}_s = 0.21 \pm 0.03\ \mbox{fm}^2 $.
A noticeable point of the pole-fit analysis is that it has the different
sign of the strange electric radius, compared with almost other models.

Another interesting approach is the kaon-loop calculation.
The main idea of the kaon-loop calculation is that the strangeness
content of the nucleon exists as a pair of $K\Lambda$ or $K\Sigma$
components.
 Koepf {\em et al.}~\cite{Koepfetal} first evaluated $\mu_s$ and 
$\langle r^2\rangle^{\rm Dirac}_s$, considering the possible kaon loops
relevant for the strange vector form factors.  
 However, Ref.~\cite{Koepfetal}
failed to include {\em seagull} terms which are 
essential to satisfy the Ward-Takahashi identity in the vector 
current sector.  Musolf {\em et al.}~\cite{musolfa} 
added these seagull terms and obtained
$\mu_s = -(0.31\rightarrow 0.40) \ \mu_N$ and 
$\langle r^2\rangle^{\rm Dirac}_s = -(6.68\rightarrow 6.90)\times10^{-3}
\ \mbox{fm}^2 $. 
The prediction of $\langle r^2\rangle^{\rm Dirac}_s$ in the kaon-loop 
calculation is found to be much smaller than the pole-fit analysis.  
To reconcile the conflict between the pole-fit analysis and the 
kaon-loop calculation, Refs.~\cite{Cohenetal,forkel} 
suggested the combination of the vector meson dominance (VMD) and
$\omega-\phi$ mixing in the vector-isovector channel 
with the kaon-loop calculation.
The value of $\langle r^2\rangle^{\rm Dirac}_s$ in Ref.~\cite{Cohenetal} 
appeared to be larger than that of the kaon-loop calculation but
still conspicuously smaller than that of the pole-fit analysis:
$\langle r^2\rangle^{\rm Dirac}_s=-(2.42\rightarrow 2.45)\times 10^{-2}
\ \mbox{fm}^2 $.  Ref.~\cite{forkel} evaluated also the
strange vector form factors and     
discussed to a great extent several different 
theoretical estimates.    

The SU(3) Skyrme model with pseudoscalar mesons~\cite{Parketal}
and with vector mesons~\cite{pw} 
estimated, respectively, $\mu_s=-0.13$, $\mu_s=-0.05$ 
and $\langle r^2\rangle^{\rm Dirac}_s=-0.10\;\mbox{fm}^2$,
$\langle r^2\rangle^{\rm Dirac}_s=0.05\;\mbox{fm}^2$.  Most recently, 
Leinweber obtained $\mu_s = -0.75\pm 0.30\;\mu_N$ 
which appears to be much larger than other models. 
    
In this paper, we aim at investigating the strange vector form factors
and related strange observables in the SU(3) 
chiral quark-soliton model ($\chi$QSM), often called semi-bosonized 
SU(3) Nambu-Jona-Lasinio model (NJL).
The model is based on the interaction of quarks with Goldstone bosons
and has been shown to be quite successful in reproducing static properties
of the baryons such as mass splitting~\cite{Blotzetal,Weigeletal},
axial constants~\cite{BlotzPraGoeke} and magnetic moments~\cite{Kim1}
and their form factors~\cite{Kim2,Kim3}.  
In a recent review~\cite{Review}, 
one can easily see how well the model describe the baryonic observables. 
In particular, since the strange vector form factors are deeply 
related to the electromagnetic form factors~\cite{jaffe,Hammeretal}
being well described in $\chi$QSM,
it is quite interesting to study them in the same framework.  

The strangeness content of the nucleon can be interpreted in
terms of the $\Lambda K$- and $\Sigma K$-components
~\cite{Koepfetal,musolfa}.  
It implies that 
one need incorporate the kaonic cloud properly in order to calculate
the strange vector form factor.  
In fact, the strange electric charge radius in a hedgehog model
is proportional to the inverse of the kaon mass, which means that
it is very sensitive to the tail of the kaonic cloud.
Hence, it is of great significance to take care that the kaonic
cloud has a proper Yukawa asymptotics
in order to evaluate the strange vector form factors rightly.
 In this respect the present paper provides a clear improvement
over the results of ref.~\cite{Review}.
In a recent study on the 
kaonic effect on the neutron electric form factor~\cite{WatKimGo},
it was shown that even the calculation of the 
neutron electric form factor requires a proper kaon tail.
In the same spirit, we expect that the kaonic cloud will 
have a decisive effect on the strange vector form factors.

The outline of the paper is as follows: Section II sketches
the general formalism for obtaining the strange vector form factors 
in the framework of $\chi$QSM.  Section III
presents the corresponding results and discuss them.
Section IV contains a summary and draws the conclusion of the present
work.
\section{General formalism}
In this section we briefly review the formalism of $\chi$QSM.
Details can be found in ref.~\cite{Review}.
We start with the low-energy partition function in Euclidean space
given by the functional integral over pseudoscalar meson ($\pi^a$) and 
quark fields($\psi$):
\begin{eqnarray}
{\cal Z} &=& \int {\cal D} \psi {\cal D} \psi^\dagger {\cal D}
\pi^a \exp{\left( -\int d^4x \psi^\dagger iD \psi\right)},  
\nonumber \\
& = & \int {\cal D} \pi^a \exp{(-S_{eff}[\pi])},
\label{Eq:action}
\end{eqnarray}
where $S_{eff}$ is the effective action 
\begin{equation}
S_{eff}[\pi] \;=\;-\mbox{Sp} \mbox{ln}iD.
\end{equation}
$iD$ represents the Dirac differential operator
\begin{equation}
iD \;=\; \beta (- i \rlap{/}{\partial} + \hat{m} + MU^{\gamma5})
\end{equation}
with the pseudoscalar chiral field
\begin{equation}
U^{\gamma_5}\;=\; \exp{(i\pi^a \lambda^a \gamma_5)}\;=\;
\frac{1+\gamma_5}{2} U + \frac{1-\gamma_5}{2}U^\dagger.
\end{equation}
$\hat{m}$ is the matrix of the current quark mass given by
\begin{equation}
\hat{m} \;=\;\mbox{diag} (m_u,m_d,m_s) 
\;=\; m_0{\bf 1} \;+\; m_8 \lambda_8,
\label{Eq:mass}
\end{equation}
where $\lambda^a$ designate the usual Gell-Mann matrices normalized as 
$\mbox{tr}(\lambda^a\lambda^b) = 2 \delta^{ab}$.  
Here, we have assumed isospin symmetry ($m_u=m_d$).  
$M$ stands for the 
dynamical quark mass arising from the spontaneous chiral
symmetry breaking, which is in general momentum-dependent~\cite{dp}. 
We regard $M$ as a constant and introduce the proper-time regularization
for convenience.  The $m_0$ and $m_8$ in Eq.~(\ref{Eq:mass}) 
are defined, respectively,
by
\begin{equation}
m_0\;=\; \frac{m_u+m_d+m_s}{3},\;\;\;\;\;m_8\;=\; 
\frac{m_u+m_d-2m_s}{2\sqrt{3}}.
\end{equation}
 The operator $i D$ is expressed in Euclidean space in terms of
the Euclidean time derivative $\partial_\tau$
and the Dirac one--particle Hamiltonian $H(U^{\gamma_5})$
\begin{equation}
i D \; = \; \partial_\tau \; + \; H(U^{\gamma_5}) 
+ \beta\hat{m} - \beta\bar{m}{\bf 1} 
\label{Eq:Dirac} 
\end{equation}
with
\begin{equation}
H(U^{\gamma_5}) \; = \; \frac{\vec{\alpha}\cdot \nabla}{i} 
\;+\; \beta MU^{\gamma_5} \;+\;\beta \bar{m}{\bf 1}.
\label{Eq:hamil}
\end{equation}
$\bar{m}$ is introduced in such a way that it produces a correct
Yukawa-type asymptotic behavior of the profile function.
$\beta$ and $\vec{\alpha}$ are the well--known 
Dirac Hermitian matrices.  
The $U$ is assumed to have a structure corresponding to the so-called
trivial embedding of the SU(2)-hedgehog into SU(3):
\begin{equation}
U\;=\; \left ( \begin{array}{cc}
U_0 & 0 \\ 0 & 1 \end{array} \right ),
\label{Eq:imbed}
\end{equation}
with
\begin{equation}
U_0\;=\;\exp{[i\vec{n}\cdot\vec{\tau}P(r)]} .
\label{Eq:profile}
\end{equation}
The profile function $P(r)$ is determined numerically by solving the
Euler-Lagrange equation corresponding to 
$\frac{\delta S_{eff}}{\delta P(r)}=0$.  
This yields a selfconsistent
classical field $U_0$ and a set of single quark energies and corresponding
states $E_n$ and $\Psi_n$.  
Note that the $E_n$ and $\Psi_n$ do not constitute
the nucleon $|N\rangle$ yet because the collective spin and and isospin
quantum numbers are missing.  
Those are obtained by the semiclassical quantization procedure,
described below in the context of the strange form factors.

The information of the strange vector form factors 
in the nucleon is contained in the quark matrix 
elements as follows:
\begin{equation}
\langle N'(p')| J^{s}_{\mu} | N(p) \rangle \;=\;
\langle N'(p')| \bar{s} \gamma_\mu s | N(p) \rangle.
\label{Eq:mat}
\end{equation}
The strange Dirac form factors of the nucleon are defined by the
matrix elements of the $J^{s}_{\mu}$:
\begin{equation}
\langle N'(p')| J^{s}_{\mu} | N(p) \rangle\;=\;
\bar{u}_{N} (p') \left[\gamma_\mu F^{s}_1 (q^2) + i \sigma_{\mu\nu}
\frac{q^\nu}{2M_N} F^{s}_2 (q^2)\right] u_N (p),
\end{equation}
where $q^2$ is the square of the four momentum transfer $q^2=-Q^2$
with $Q^2>0$.  $M_N$ and $u_N(p)$ stand for the nucleon mass and
its spinor, respectively.
The strange quark current $J^{s}_{\mu}$ can be expressed in terms
of the baryon current and the hypercharge current:
\begin{equation}
J^{s}_{\mu}\;=\; \bar{s} \gamma_{\mu} s 
\;=\; J^{B}_{\mu} - J^{Y}_{\mu}
\;=\;\bar{q}\gamma_\mu\hat{Q}_s q,
\label{Eq:scur}
\end{equation}
where
\begin{eqnarray}
J^{B}_{\mu} & = & \frac{1}{N_c} \bar{q} \gamma_\mu q, \nonumber \\
J^{Y}_{\mu} & = & \frac{1}{\sqrt{3}} 
\bar{q} \gamma_\mu \lambda_8 q \nonumber \\
\hat{Q}_s & = &\frac{1}{N_c}
-\frac{1}{\sqrt{3}}\lambda_8,
\end{eqnarray}
where $N_c$ denotes the number of colors of the quark. 
$\hat{Q}_s=\mbox{diag}(0,0,1)$ is called {\em strangeness operator}:
We employ the non-standard sign convention
used by Jaffe~\cite{jaffe} for the strange current.
The baryon and hypercharge currents are equal to
the singlet and octet currents, respectively. 

The strange Dirac form factors $F^{s}_1$ and $F^{s}_{2}$ 
can be written in terms of the strange Sachs form factors, 
$G^{s}_E(Q^2)$ and $G^{s}_M(Q^2)$:
\begin{eqnarray}
G^{s}_E (Q^2) &=& F^{s}_1 (Q^2) - \frac{Q^2}{4M^2_{N}} F^{s}_{2}(Q^2)
\nonumber \\
G^{s}_M (Q^2) &=& F^{s}_1 (Q^2) + F^{s}_{2}(Q^2).
\end{eqnarray}
In the non--relativistic limit($Q^2 \ll M^{2}_N$), 
the Sachs-type form factors $G^{s}_{E}(Q^2)$ and $G^{s}_{M} (Q^2)$
are related to the time and space components of the strange
current, respectively:
\begin{eqnarray}
\langle N'(p') |J^{s}_{0}(0) | N(p) \rangle & = &  
G^{s}_{E} (Q^2) \nonumber \\
\langle N'(p') | J^{s}_{i}(0) | N(p) \rangle & = &  
\frac{1}{2M_N} G^{s}_{M} (Q^2) i 
\epsilon_{ijk} q^j \langle \lambda'| \sigma_k | \lambda \rangle,
\label{Eq:gm}
\end{eqnarray}
where $\sigma_k$ stand for Pauli spin matrices.  The $| \lambda\rangle$
is the corresponding spin state of the nucleon.
The matrix elements of the strange quark current can be related to
a correlator:
\begin{equation}
\langle N'(p')| \bar{s}\gamma_{\mu} s | N(p) \rangle 
\smash{\mathop{\sim}\limits_{T\rightarrow \infty}}
\langle 0 | J_{N'} (\vec{x},T/2) \bar{q} \gamma_{\mu} \hat{Q}_s q 
J^{\dagger}_{N} (\vec{y},-T/2) |0 \rangle.
\label{Eq:expect}
\end{equation}    
The nucleon current $J_N$ can be built from $N_c$ quark fields 
\begin{equation}
J_N(x)\;=\; \frac{1}{N_c !} \epsilon_{i_{1} \cdots i_{N_c}} 
\Gamma^{\alpha_1 \cdots
\alpha_{N_c}}_{JJ_3TT_3Y}\psi_{\alpha_1i_1}(x)
\cdots \psi_{\alpha_{N_c}i_{N_c}}(x).
\label{Eq:corr}
\end{equation}
$\alpha_1 \cdots\alpha_{N_c}$ denote spin--flavor indices, while
$i_1 \cdots i_{N_c}$ designate color indices.  The matrices 
$\Gamma^{\alpha_1 \cdots\alpha_{N_c}}_{JJ_3TT_3Y}$ are taken to endow 
the corresponding current with the quantum numbers $JJ_3TT_3Y$.
In our model, Eq. (\ref{Eq:expect}) is represented 
by the Euclidean functional integral with regard to quark 
and pseudo-Goldstone fields:
\begin{eqnarray}
\langle N'(p')| \bar{q}\gamma_{\mu} \hat{Q}_s q | N(p) \rangle
& = & \frac{1}{\cal Z} \lim_{T\rightarrow \infty} 
\exp{(ip_4 \frac{T}{2}
- ip'_{4} \frac{T}{2})} \nonumber \\
& \times & \int d^3 x d^3 y 
\exp{(-i \vec{p'} \cdot \vec{y} + i \vec{p} \cdot \vec{x})} 
\int {\cal D}U \int {\cal D} \psi \int {\cal D}\psi^\dagger 
\nonumber \\
& \times & \; J_{N'}(\vec{y},T/2)q^\dagger(0) 
\beta \gamma_\mu \hat{Q}_s q(0) J^{\dagger}_{N} (\vec{x}, -T/2) 
\nonumber \\ & \times &
\exp{\left[ - \int d^4 z \psi^\dagger i D \psi \right ]},
\label{Eq:stff1}
\end{eqnarray}
where ${\cal Z}$ stands for the normalization factor
which is expressed by the same functional integral but without
the quark current operator $\bar{s} \gamma_\mu s$.  
Eq.(\ref{Eq:stff1}) can be decomposed into valence and sea contributions:
\begin{equation}
\langle N'(p')| \bar{q}\gamma_{\mu} \hat{Q}_s q | N(p) \rangle\;=\;
\langle N'(p')| \bar{q}\gamma_{\mu} \hat{Q}_s q | N(p) \rangle_{val}\; +\; 
\langle N'(p')| \bar{q}\gamma_{\mu} \hat{Q}_s q | N(p) \rangle_{sea},
\end{equation}
where 
\begin{eqnarray}
\langle N'(p')| V_\mu(0) | N(p) \rangle_{val} & = & 
\frac{1}{\cal Z} 
\Gamma^{\beta_1 \cdots \beta_{N_c}}_{J'J'_3T'T'_3Y'}
\Gamma^{\alpha_1 \cdots \alpha_{N_c}*}_{JJ_3TT_3Y} 
\lim_{T \rightarrow \infty} \exp{(ip_4 \frac{T}{2}
- ip'_{4} \frac{T}{2})}
\nonumber \\
& \times & \int d^3x d^3y \exp{(-i\vec{p'}\cdot\vec{y}
+ i \vec{p} \cdot \vec{x})} \nonumber \\
& \times & \int {\cal D}U \exp{(-S_{eff})} 
\sum^{N_c}_{i=1} \; _{\beta_i}\langle\vec{y}, {\mbox T}/2|
\frac{1}{i D} | 0,t_z \rangle_{\gamma} 
[\beta \gamma_\mu \hat{Q}_s ]_{\gamma \gamma'} \nonumber \\ & \times &
_{\gamma'}\langle 0,t_z | \frac{1}{i D} 
| \vec{x}, -{\mbox T}/2\rangle_{\alpha_i} 
\prod^{N_c}_{j \neq i}
\; _{\beta_j} \langle \vec{y},{\mbox T}/2 |
\frac{1}{i D} |\vec{x}, -{\mbox T}/2 \rangle_{\alpha_j}
\label{Eq:val1}
\end{eqnarray}
and 
\begin{eqnarray}
\langle N'(p')| V_\mu(0) |N(p) \rangle_{sea} & = & 
-\frac{N_c}{\cal Z} 
\Gamma^{\beta_1 \cdots \beta_{N_c}}_{J'J'_3T'T'_3Y'}
\Gamma^{\alpha_1 \cdots \alpha_{N_c}*}_{JJ_3TT_3Y} 
\lim_{T \rightarrow \infty} \exp{(ip_4 \frac{T}{2}
- ip'_{4} \frac{T}{2})}
\nonumber \\
& \times & \int d^3x d^3y \exp{(-i\vec{p'}\cdot\vec{y}
+ i \vec{p} \cdot \vec{x})} \nonumber \\
& \times & \int {\cal D}U \exp{(-S_{eff})}
{\rm Tr}\ _{\gamma \lambda} 
\langle 0, t_z | \frac{1}{i D}[\beta \gamma_\mu] 
\hat{Q}_s | 0, t_z \rangle        \nonumber \\ 
& \times & \prod^{N_c}_{i=1}\; _{\beta_i} 
\langle \vec{y},{\mbox T}/2 |   \frac{1}{i D} 
| \vec{x}, -{\mbox T}/2\rangle_{\alpha_i}.
\label{Eq:sea1}
\end{eqnarray}
$S_{eff}$ is the effective chiral action expressed by
\begin{equation}
S_{eff} \;=\; -N_c {\mbox{Sp}} \mbox{ln}\left 
[ \partial_\tau \;+\; H(U^{\gamma_5})
\; + \; \beta \hat{m} \;-\;\beta\bar{m}{\bf 1}\right ].
\end{equation}

 In order to perform the collective quantization,
we have to integrate Eqs.~(\ref{Eq:val1}) and (\ref{Eq:sea1}) over 
small oscillations of the pseudo-Goldstone field around
the saddle point Eq.~(\ref{Eq:imbed}).  This will not be done
except for the zero modes.  The corresponding fluctuations of the
pion fields are not small and hence cannot be neglected.
 The zero modes are relevant to continuous symmetries
in our problem.  In particular,   
we have to take into account the translational
zero modes properly in order to evaluate form factors,  
since the soliton is not invariant under translation and its translational
invariance is restored only after integrating over the translational
zero modes.  Explicitly, the zero modes are taken into account by 
considering a slowly {\em rotating} and {\em translating} hedgehog:
\begin{equation}
\tilde{U}(\vec{x}, t)\;=\; A(t) 
U(\vec{x}-\vec{Z}(t)) A^{\dagger} (t).
\label{Eq:rot}
\end{equation}
$A(t)$ belongs to an SU(3) unitary matrix.
 The Dirac operator $i\tilde{D}$ in Eq.~(\ref{Eq:Dirac}) can be written
as
\begin{equation} 
i \tilde{D} \; = \; \left(\partial_\tau \; + \; H(U^{\gamma_5}) 
\;+\;A^{\dagger} (t) \dot{A}(t)
\;-\; i \beta \dot{\vec{Z}} \cdot \nabla
\;+\; \beta A^{\dagger} (t) (\hat{m}-\bar{m}{\bf 1}) A(t) \right).
\end{equation}
The corresponding collective action is expressed by
\begin{eqnarray}
\tilde{S}_{eff} & = & -N_c {\rm Sp}\
\mbox{ln}\left [ \partial_\tau \;+\; H(U^{\gamma_5}) 
\;+\; A^{\dagger} (t) \dot{A}(t)
\;-\;  i \beta \dot{\vec{Z}} \cdot \nabla \right .
\nonumber \\  
&  & \left .
\; + \; \beta A^{\dagger}(t) (\hat{m}-\bar{m}{\bf 1}) A(t) 
\;-\; \beta A^{\dagger}(t) s_\mu \gamma_\mu \hat{Q}_s A(t) \right ] 
\label{Eq:effact}
\end{eqnarray}
with the angular velocity 
\begin{equation}
A^{\dagger}(t)\dot{A}(t) \;=\; i\Omega_E \;=\; 
\frac{1}{2} i \Omega^{a}_{E} \lambda^a 
\end{equation}
and the velocity of the translational motion
\begin{equation}
\dot{\vec{Z}}\;=\; \frac{d}{dt} \vec{Z} .
\end{equation} 
  
Hence, Eq.~(\ref{Eq:val1}) 
and Eq.~(\ref{Eq:sea1}) can be written in terms of
the rotated Dirac operator $i\tilde{D}$ and chiral effective action
$\tilde{S}_{eff}$.     
 The functional integral over the pseudoscalar field
$U$ is replaced by the path integral which can be calculated
in terms of the eigenstates of the Hamiltonian 
corresponding to the collective action 
and these Hamiltonians can be diagonalized in an exact manner.

We take into account the rotational $1/N_c$ and
$m_s$ corrections up to linear order: 
\begin{equation}
\frac{1}{i\tilde{D}}\simeq \frac{1}{\partial_\tau + H}
+\frac{1}{\partial_\tau + H}(-i\Omega_E)\frac{1}{\partial_\tau + H}
+\frac{1}{\partial_\tau + H}
(-\beta A^\dagger [\hat{m}-\bar{m}{\bf 1}]A) \frac{1}{\partial_\tau + H}.
\end{equation}
When the mass corrections are considered, SU(3) symmetry is no more
exact.  Thus, the eigenfunctions of the collective Hamiltonian are 
neither in a pure octet nor in a pure decuplet 
but in mixed states with higher representations:
\begin{equation}
| 8, N \rangle \;=\; | 8,N \rangle \;+ \; 
c_{\bar{10}} | \bar{10},N \rangle
\;+\;c_{27} | 27,N \rangle
\label{Eq:wfc}
\end{equation}
with
\begin{equation}
c_{\bar{10}} \;=\; \frac{\sqrt{5}}{15}(\sigma - r_1)I_2 m_s,
\;\; c_{27} \;=\; \frac{\sqrt{6}}{75}(3\sigma + r_1 - 4r_2)
I_2 m_s.
\label{Eq:g2}
\end{equation}
The constant $\sigma$ is related to the SU(2) $\pi N$ sigma term 
$\Sigma_{SU(2)}\;=\;3/2 (m_u + m_d) \sigma$ and $r_i$ designates
$K_i/I_i$, where $K_i$ stand for the anomalous moments of inertia
defined in~Ref.\ \cite{Blotzetal}.

Having carried out a lengthy manipulation 
(for details, see Ref.\cite{Review}), 
we arrive at our final expressions for the strange vector form factors.  
The Sachs strange electric form factor $G^{s}_{E}$ is expressed as follows
 (see appendix A for detail):
\begin{eqnarray}
G^{s}_{E}(\vec{Q}^2)& = & \left (1 -
\langle D^{(8)}_{88}\rangle_N \right) {\cal B} (Q^2)
\nonumber \\
&+& \langle D^{(8)}_{8a} J_a\rangle_N 
\frac{2{\cal I}_1 (Q^2)}{\sqrt{3} I_1}    
+ \langle D^{(8)}_{8p} J_p\rangle_N 
\frac{2{\cal I}_2 (Q^2)}{\sqrt{3} I_2}    
 \nonumber \\
& + &  3 \left(m_0 -\bar{m} + \frac{m_8}{\sqrt{3}}
\langle D^{(8)}_{88}\rangle_N\right) {\cal C} (Q^2)
\nonumber \\
& - & 
\langle D^{(8)}_{8a} D^{(8)}_{8a}\rangle_N
\frac{4 m_8}{\sqrt{3}I_1} \left (I_1 {\cal K}_1 (\vec{Q}^2)
-{\cal I}_1 (\vec{Q}^2) K_1\right ) \nonumber \\
& - & 
\langle D^{(8)}_{8p} D^{(8)}_{8p}\rangle_N
\frac{4 m_8}{\sqrt{3}I_2} \left (I_2 {\cal K}_2 (\vec{Q}^2)
-{\cal I}_2 (\vec{Q}^2) K_2 \right ),
\label{Eq:elecf}  
\end{eqnarray}
 $I_i$ and $K_i$ are the moments of inertia and anomalous
moments of inertia~\cite{Blotzetal}, respectively,
${\cal B}$, ${\cal I}_i$, and ${\cal K}_i$
correspond to the baryon number, moments of inertia,
and the anomalous moments of inertia
at $Q^2=0$, respectively.  
From Eq.(\ref{Eq:elecf}), we can easily see that at $Q^2=0$ the strange
electric form factor $G^{s}_{E}$ vanishes (note that 
${\cal C}(Q^2=0)=0$).  Making use of the relation 
$\sum^{8}_{a=1}D^{(8)}_{8a} J_a = -\sqrt{3}Y/2$ and
$J_8 = -N_c /(2\sqrt{3})$, we obtain
$G^{s}_{E}(Q^2=0)=B-Y=S$.  Since the net strangeness
of the nucleon is zero, $G^{s}_{E}$ at $Q^2=0$ must
vanish.  
The final expression of the Sachs strange magnetic form factor is
written (see appendix A for detail) by
\begin{eqnarray}
G^{s}_{M} (\vec{Q}^2) & = & \frac{M_N}{|\vec{Q}|} 
\left [-\frac{\langle D^{(8)}_{83}\rangle_N}{\sqrt{3}} 
\left({\cal Q}_0 (\vec{Q}^2) 
\; +\; \frac{{\cal Q}_1(\vec{Q}^2)}{I_1}
\; +\; \frac{{\cal Q}_2(\vec{Q}^2)}{I_2}\right) \right.
\nonumber \\
&+& \langle (D^{(8)}_{88}-1) J_3\rangle_N  
\frac{{\cal X}_1 (\vec{Q}^2)}{3 I_1} 
\;+\;   \langle d_{3pq}D^{(8)}_{8p}J_q \rangle_N \delta_{pq}
\frac{{\cal X}_2 (\vec{Q}^2)}{\sqrt{3} I_2} 
\nonumber \\
&+& 6 (m_0 - \bar{m}) \langle D^{(8)}_{83}\rangle_N
{\cal M}_0 (\vec{Q}^2)
\;+\;2 \sqrt{3} m_8 \langle D^{(8)}_{88}D^{(8)}_{83} \rangle_N 
{\cal M}_0 (\vec{Q}^2) \nonumber \\
& + &  m_0  \langle D^{(8)}_{83} \rangle_N
\left(2 {\cal M}_1 (\vec{Q}^2)  \;-\; 
\frac{2}{\sqrt{3}} r_1 {\cal X}_1 (\vec{Q}^2)  \right) 
\nonumber \\
& + &  m_8 \langle D^{(8)}_{83} D^{(8)}_{88} \rangle_N
\left(2 {\cal M}_1 (\vec{Q}^2)  \;-\; 
\frac{2}{3} r_1 {\cal X}_1 (\vec{Q}^2)  \right) 
\nonumber \\
& + & \left . \sqrt{3} m_8   
\langle d_{3pq}D^{(8)}_{8p}D^{(8)}_{8q} \rangle_N\delta_{pq}
\left(2 {\cal M}_2 (\vec{Q}^2) 
\; - \;  
\frac{2}{3} r_2 {\cal X}_2 (\vec{Q}^2)  \right) \right ],
\label{Eq:magf}
\end{eqnarray}
\section{Results and Discussions}
In order to evaluate Eqs. (\ref{Eq:elecf},\ref{Eq:magf})
numerically, we follow the Kahana-Ripka discretized basis 
method~\cite{kr}.
However, note that it is of great importance to use a reasonably
large size of the box ($D\approx 10 \;\mbox{fm}$) so as to get
a numerically stable results.    
The present SU(3) $\chi$QSM (equivalent to SU(3) NJL on the chiral 
circle) contains four free parameters.  Two of them are fixed in
the meson sector by adjusting them to the pion mass, 
$m_\pi=139\ \mbox{MeV}$, the pion decay constant, 
$f_\pi=93\ \mbox{MeV}$, and the kaon mass, $m_{\rm K}=496 \ \mbox{MeV}$.  
As for the fourth parameter, {\em i.e.} the constituent mass $M$
of up and down quarks, values around $M=420\ \mbox{MeV}$ have been used
because they have turned out to be the most appropriate one for the
description of nucleon observables and form factors 
(see ref.~\cite{Review}).  In fact, $M=420\ \mbox{MeV}$ is the
preferred value, which is always used in this paper.
For the description of the baryon sector, we choose the method of Blotz 
{\em et al.}~\cite{Blotzetal} modified for a finite meson mass. 
The resulting strange current quark mass comes out around 
$m_s=180\ \mbox{MeV}$.  In order to illustrate the effect of the $m_s$
the calculations in the baryonic sector are performed with both
$m_s=0$ and finite $m_s$.  One should note that
a SU(3)-calculation with $m_s=0$ does not correspond to a SU(2) 
calculation, since the spaces, in which the collective
quantization are performed, are different.  

In the present calculation, the mass parameter $\bar{m}$ plays
a pivotal role, because it makes the solitonic profile $P(r)$ of
Eq.(\ref{Eq:profile}) incorporate the
proper Yukawa-type tail:
\begin{equation}
P(r) \sim  \exp{\left(-\mu r\right)} 
\frac{1 + \mu r}{r^2},
\end{equation}
where $\mu$ denotes the meson mass suppressing the tail of
the profile.  In fact, $\mu$ is related to the $\bar{m}$
in a non-linear way whose details can be extracted from the meson 
expansion of ref.~\cite{WatKimGo}.  In the end it turns out that
when $\bar{m} = (m_u+m_d)/2$, the $\mu$ becomes the pion mass
$m_\pi=139$ MeV, while $\mu$ corresponds to 
the kaon mass $m_{\rm K} \simeq 490$ MeV for the $\bar{m} \simeq 75$ MeV.  
Actually, since the hedgehog formalism forces us to have just one profile
function and hence all mesonic fields to have the same
Yukawa mass in the tail, one has to decide if one wants the pion
tail {\em or} the kaon tail to be correct.  For previous investigations
of electromagnetic properties of the nucleon it was preferable
to have a correct pion tail at the expense of a poor kaon tail
and in this sense all calculations of ref.~\cite{Review} 
have been performed.  For the present investigation of the strange
vector form factors it is more desirable to have a kaon tail
with a Yukawa mass, $\mu$, being equal to the kaon mass, $m_{\rm K}$,
and hence we prefer in this paper $\bar{m}\simeq 75$ MeV corresponding 
to $\mu = m_{\rm K}\simeq 490$ MeV.  We know that the pion tail is
now too short.  However, we do not expect it to matter for the
strange vector form factors.  We give the results with $\mu=m_{\pi}=
139$ MeV for comparison.
Figure 1 shows the strange electric form factor $G^{s}_{E} (Q^2)$, 
as the constituent quark mass $M$ is varied from 400 MeV to 450 MeV
with $m_s = 180$ MeV.  
The strange electric form factor $G^{s}_{E}$ decreases as
$M$ increases.  
Fig. 2 displays the effect of the $m_s$ corrections
on the $G^{s}_{E}$.  When they are turned off,
the $G^{s}_{E}$ becomes negative.  At first glance,  
it seems surprising, compared to the results with $\mu = m_\pi$~\cite{Review},
though in Ref.~\cite{Review} the $m_s$ corrections turn out to be
very large.  However, replacing $\mu = m_\pi$ by $\mu = m_{\rm K}$ 
leads to the fact that the leading-order contribution and rotational
$1/N_c$ corrections are sizably reduced, while the $m_s$ corrections
are relatively not much weakened.  As a result, the $m_s$ corrections
change the sign of the $G^{s}_{E}$ as shown in Fig. 2.
This raises the question whether higher than first order corrections
in $m_s$ are important.  This question will be investigated in
near future.  

In Fig. 3 the effect of the kaonic cloud is well explained.  
The $G^{s}_{E}$ with $\mu = m_{\rm K}$ is almost three times
smaller than that with $\mu = m_\pi$.  This remarkable result is
in line with the recent investigation of the kaonic effects
on the neutron electric form factor~\cite{WatKimGo}.
These kaonic effects can be understood more explicitly 
by evaluating the strange electric radii.

The Sachs and Dirac mean-square strange radii are, respectively, defined
by
\begin{equation}
\left. \langle r^2\rangle^{\rm Sachs}_{s} \;=\; 
-6\frac{d G_{E}^{s} (Q^2)}{dQ^2} 
\right|_{Q^2=0}  ,\;\;\;\;
\left. \langle r^2\rangle^{\rm Dirac}_{s} \;=\; 
-6\frac{d F_{1}^{s} (Q^2)}{dQ^2} 
\right|_{Q^2=0}
\end{equation}
We obtain  
$\left.\langle r^2\rangle^{\rm Sachs}_{s}\right|_{\mu = m_\pi}
=-3.5\times 10^{-1} \;\mbox{fm}^2$
and $\left. \langle r^2\rangle^{\rm Dirac}_{s}\right |_{\mu = m_{\rm K}}
=-3.2\times 10^{-1} \;\mbox{fm}^2$
with the pion tail, whereas we get 
$\left. \langle r^2\rangle^{\rm Sachs}_{s}\right|_{\mu = m_{\rm K}}
=-9.5 \times 10^{-2}\;\mbox{fm}^2$
and $\left. \langle r^2\rangle^{\rm Dirac}_{s}\right|_{\mu = m_{\rm K}}
=-5.0 \times 10^{-2} \;\mbox{fm}^2$ .  
Again, we find that the case of $\mu = m_{\rm K}$ is three times smaller
than that of $\mu = m_\pi$.
The mean-square strange radii depend on the meson mass $\mu$
which suppresses the tail of the profile, {\em i.e.} 
$\langle r^2\rangle^{\rm Sachs}_{s} \sim 1/\mu$.
From such a behavior of the $\langle r^2\rangle^{\rm Sachs}_{s}$, 
we can derive the relation
\begin{equation}
\frac{\left. \langle r^2\rangle^{\rm Sachs}_{s} \right |_{\mu=m_\pi}}
{\left. \langle r^2\rangle^{\rm Sachs}_{s} \right |_{\mu=m_{\rm K}}}
= \frac{m_{\rm K}}{m_\pi} \simeq 3.5 .
\label{Eq:radius}
\end{equation}  
Eq.(\ref{Eq:radius}) explains the decrease of the 
$\langle r^2\rangle_{s}$ with $\mu=m_{\rm K}$.  

Fig. 4 illustrates the strange electric densities weighted with $r^2$.
As expected from the above discussion, the kaonic cloud diminishes
the strange electric density sizably.  

Fig. 5 draws the strange magnetic form factor.  In contrast to the
$G^{s}_{E}$, the $G^{s}_{M}$ increases slowly with the increasing
constituent quark mass apart from the small $Q^2$ region (
below about $Q^2=0.2\;\mbox{GeV}^2$).  Fig. 6 shows that the 
$m_s$ corrections has a small effect on the $G^{s}_{M}$.
It is very different from what we saw in the case of the $G^{s}_{M}$.
However, Eq.(\ref{Eq:magf}) explains why the $m_s$ corrections
are reduced by $\mu = m_{\rm K}$: The fourth term including 
$(m_0 - \bar{m})$ becomes smaller on account of the large $\bar{m}$.

Fig. 7 displays the effect of the kaonic cloud on the $G^{s}_{M}$.
With $\mu$ increased to be $m_{\rm K}$, we find that the $G^{s}_{M}$
is almost 50 $\%$ enhanced.  
Fig. 8 draws the corresponding magnetic densities weighted by $r^2$.  

In table 1, the strange magnetic moments $\mu_s$ and
mean-square strange radii $\langle r^2\rangle_{s}$
are displayed as a function of $M$
and $m_s$ in the case of $\mu = m_{\rm K}$, while in table 2
the same in the case of $\mu = m_\pi$.  
According to our philosophy the values of table 1 with $\mu = m_{\rm K}$
are the results of the present model.  If one fixes the constituent quark
mass $M$ to a value of $M=420$ MeV then other 
baryonic properties such as the octet-decuplet mass splitting
and various form factors are reproduced as well.  
Hence, we have $M=420$ MeV as canonical value.
In table 3, we have made a comparison for the $\mu_s$ and 
$\langle r^2\rangle^{\rm Sachs}_{s}$ between different models.

We want to take the occasion to comment on Ref.~\cite{tueb}
which provides calculations for a correct pion tail and poor kaon tail,
$\mu=m_{\pi}$.
Though Ref.~\cite{tueb} seems to use in this case 
the same model as the present
work, there are significant differences between
these two papers.  First, Weigel {\em et al.}~\cite{tueb} do not
consider rotational $1/N_c$ corrections in contrast to the present paper.
This has the immediate consequence that the magnetic moments of 
Weigel {\em et al.} are $\mu_p = 1.06\;\mu_N,\;
\mu_n=-0.69\;\mu_N$ for the nucleon
\footnote{For this comparison, the constituent quark mass
$M=450\;\mbox{MeV}$ is chosen.  In case of $M=420\;\mbox{MeV}$,
we have obtained $\mu_p=2.39\;\mu_N$ and 
$\mu_n=-1.76\;\mu_N$~\cite{Kim1}.}
whereas the present work (including those corrections) yields
$\mu_p = 2.20\;\mu_N,\;\mu_n=-1.59\;\mu_N$ with a far better comparison with
experiment ($\mu_p = 2.79\;\mu_N,\;\mu_n=-1.91\;\mu_N$).  Furthermore,
Weigel {\em et al.} regularize, besides the real part of the action,
also the imaginary one.  This meets problems in producing the anomaly
structure and is hence avoided in the approach of the present work.
In addition the calculation of Weigel {\em et al.} are not fully
self-consistent but use some scaling approximations.

\section{Summary and Conclusion}
In summary, we have calculated in the SU(3) chiral quark-soliton model
($\chi$QSM) often called the semibosonized SU(3) 
Nambu--Jona-Lasinio model, the strange electric and magnetic
form factors of the nucleon, $G^{s}_{E}$ and $G^{s}_{M}$
including the strange magnetic moment $\mu_s$, and the mean-square 
strange radius $\langle r^2\rangle_s$.  
The theory takes into account rotational $1/N_c$ corrections
and linear $m_s$ corrections.
Choosing the parameters of the model in such a way that the 
kaon cloud falls off with a Yukawa mass equal to the kaon mass,
we have obtained $\mu_s=-0.68 \ \mu_N$,
$\langle r^2\rangle^{\rm Dirac}_s = -0.051\ \mbox{fm}^2$ and
$\langle r^2\rangle^{\rm Sachs}_s = -0.095\ \mbox{fm}^2$.
 The results have been compared with different other models.

There are several points where the present calculations leave room for
further studies.  Apparently the dependence of the form factors
on the value of $m_s$ is quite noticeable and probably one has
to go to higher orders in perturbation theory in $m_s$.  
 Besides the strange vector form factors the strange
axial form factors are also of great interest.  Presently
we are performing investigations to clarify these questions.

\section*{Acknowledgment}
We would like to thank Chr.V. Christov, P.V. Pobylitsa,
M.V. Polyakov and W. Broniowski
for fruitful discussions and critical comments.
This work has partly been supported by the BMBF, the DFG
and the COSY--Project (J\" ulich).
\begin{appendix}
\section{}
In this appendix, we present all formulae appearing in 
Eqs.(\ref{Eq:elecf},\ref{Eq:magf}).    
\begin{eqnarray}
{\cal B}(\vec{Q}^2) & = & \int d^3 x \; j_0 (Qr)
\left [ \Psi^{\dagger}_{val}(x) \Psi_{val} (x) \;-\;\frac{1}{2} \sum_n
{\rm sgn} (E_n) \Psi^{\dagger}_{n}(x) 
\Psi_{n} (x) \right ], \nonumber \\ 
{\cal C}(Q^2) & = & -\frac{2N_c}{3} \sum_{nm}\int d^3 x j_0 (Qr)
\int d^3 y \left[\frac{\Psi^\dagger (y) 
\beta \Psi_{val} (y) \Psi^{\dagger}_{val} (x)
\Psi_n (x)}{E_n - E_{val}}   \right. \nonumber \\
& & \left. \hspace{4.5cm} \;+\;\frac{1}{2} R_{\cal M} (E_n, E_m)
\Psi^{\dagger}_{n}(y) \beta \Psi_m (y) \Psi^{\dagger}_{m} (x)
\Psi_n (x)  \right], \\
{\cal I}_1 (\vec{Q}^2) & = & \frac{N_c}{6} \sum_{n, m}
\int d^3 x \;j_0 (Qr) \int d^3 y 
\left [\frac{\Psi^{\dagger}_{n} (x) \vec{\tau} \Psi_{val} (x) \cdot 
\Psi^{\dagger}_{val} (y) \vec{\tau} \Psi_{n} (y)}
{E_n - E_{val}} \right .
\nonumber \\  & & \hspace{3cm} \;+\; \left . \frac{1}{2}
\Psi^{\dagger}_{n} (x) \vec{\tau} \Psi_{m} (x) \cdot 
\Psi^{\dagger}_{m} (y) \vec{\tau} \Psi_{n} (y) 
{\cal R}_{\cal I} (E_n, E_m) \right ],
\nonumber \\
{\cal I}_2 (\vec{Q}^2) & = &\frac{N_c}{6} \sum_{n, m^{0}}
\int d^3 x \;j_0 (Qr)\int d^3 y 
\left [\frac{\Psi^{\dagger}_{m^{0}} (x) \Psi_{val} (x) 
\Psi^{\dagger}_{val} (y) \Psi_{m{^0}} (y)}
{E_{m^{0}} - E_{val}} \right .
\nonumber \\  & & \hspace{3cm} \;+\;\left . \frac{1}{2}
\Psi^{\dagger}_{n} (x) \Psi_{m^{0}} (x) 
\Psi^{\dagger}_{m^{0}} (y) \Psi_{n} (y) 
{\cal R}_{\cal I} (E_n, E_m^{0}) \right ],
\nonumber \\
{\cal K}_1 (\vec{Q}^2) & = & \frac{N_c}{6} \sum_{n, m}
\int d^3 x \;j_0 (Qr) \int d^3 y 
\left [\frac{\Psi^{\dagger}_{n} (x) \vec{\tau} \Psi_{val} (x) \cdot 
\Psi^{\dagger}_{val} (y) \beta \vec{\tau} \Psi_{n} (y)}
{E_n - E_{val}} \right .  
\nonumber \\  & & \hspace{3cm} \;+\; \left . \frac{1}{2}
\Psi^{\dagger}_{n} (x) \vec{\tau} \Psi_{m} (x) \cdot 
\Psi^{\dagger}_{m} (y) \beta \vec{\tau} \Psi_{n} (y) 
{\cal R}_{\cal M} (E_n, E_m) 
\right ],
\nonumber \\
{\cal K}_2 (\vec{Q}^2) & = & \frac{N_c}{6} \sum_{n, m^{0}}
\int d^3 x \;j_0 (Qr) \int d^3 y 
\left [\frac{\Psi^{\dagger}_{m^{0}} (x) \Psi_{val} (x) 
\Psi^{\dagger}_{val} (y) \beta \Psi_{m{^0}} (y)}
{E_{m^{0}} - E_{val}} \right .
\nonumber \\  & & \hspace{3cm} \;+\;\left . \frac{1}{2}
\Psi^{\dagger}_{n} (x) \Psi_{m^{0}} (x) 
\Psi^{\dagger}_{m^{0}} (y) \beta \Psi_{n} (y) 
{\cal R}_{\cal M} (E_n, E_m^{0}) \right ] 
\end{eqnarray}
with regularization functions
\begin{eqnarray}
{\cal R}_{I} (E_n, E_m) & = & - \frac{1}{2\sqrt{\pi}}
\int^{\infty}_{0} \frac{du}{\sqrt{u}} \phi (u;\Lambda_i) 
\left [ \frac{E_n e^{-u E^{2}_{n}} +  E_m e^{-u E^{2}_{m}}}
{E_n + E_m} \;+\; \frac{e^{-u E^{2}_{n}} - e^{-u E^{2}_{m}}}
{u(E^{2}_{n} - E^{2}_{m})} \right ],
\nonumber \\
{\cal R}_{\cal M} (E_n, E_m) & = &
\frac{1}{2}  \frac{ {\rm sgn} (E_n) 
- {\rm sgn} (E_m)}{E_n - E_m}.
\label{Eq:regul}
\end{eqnarray}

\begin{eqnarray}
{\cal Q}_0 (\vec{Q}^2) & = & 
{N_c}\int d^3 x j_1 (qr) 
\left[ \Psi ^{\dagger}_{val}(x) \gamma_{5} 
\{\hat{r} \times \vec{\sigma} \} \cdot \vec{\tau} 
\Psi_{val} (x) \right. \nonumber \\ & & 
\left . \hspace{1cm} \;-\; 
\frac{1}{2}  \sum_n {\rm sgn} (E_n) 
\Psi ^{\dagger}_{n}(x) \gamma_{5} 
\{\hat{r} \times \vec{\sigma} \} \cdot \vec{\tau} 
\Psi_{n}(x) {\cal R}(E_n)\right ],  
\nonumber \\
{\cal Q}_1 (\vec{Q}^2) & = &  \frac{iN_c}{2}\sum_{n} 
\int d^3 x j_1 (qr)
\int d^3 y \nonumber \\  & \times &
\left[{\rm sgn} (E_n)
\frac{\Psi^{\dagger}_{n} (x) \gamma_{5}
\{\hat{r} \times \vec{\sigma} \} \times \vec{\tau}   
\Psi_{val} (x) \cdot 
\Psi^{\dagger}_{val} (y) \vec{\tau} \Psi_{n} (y)}
{E_n - E_{val}} \right .
\nonumber \\  & & 
\;+\; \left . \frac{1}{2} \sum_{m}
\Psi^{\dagger}_{n} (x)\gamma_{5} \{\hat{r} \times \vec{\sigma} \} 
\times \vec{\tau}  \Psi_{m} (x) \cdot 
\Psi^{\dagger}_{m} (y) \vec{\tau} \Psi_{n} (y) 
{\cal R}_{\cal Q} (E_n, E_m) \right ],
\nonumber \\
{\cal Q}_2 (\vec{Q}^2) & = &  \frac{N_c}{2} \sum_{m^0} 
\int d^3 x j_1 (qr)
\int d^3 y \nonumber \\  & \times &
\left[ {\rm sgn} (E_{m^0})
\frac{\Psi^{\dagger}_{m^0} (x) \gamma_{5}
\{\hat{r} \times \vec{\sigma} \} \cdot \vec{\tau}   
\Psi_{val} (x) 
\Psi^{\dagger}_{val} (y) \Psi_{m^0} (y)}
{E_{m^0} - E_{val}} \right .
\nonumber \\  & + & 
\left . \sum_{n}
\Psi^{\dagger}_{n} (x)\gamma_{5} \{\hat{r} \times \vec{\sigma} \} 
\cdot \vec{\tau}  \Psi_{m^0} (x) 
\Psi^{\dagger}_{m^0} (y) \Psi_{n} (y) 
{\cal R}_{\cal Q} (E_n, E_{m^0}) \right ],
\nonumber \\
{\cal X}_1 (\vec{Q}^2) & = & N_c
\sum_{n} \int d^3 x j_1 (qr)
\int d^3 y \left[
\frac{\Psi^{\dagger}_{n} (x)\gamma_{5}
\{\hat{r} \times \vec{\sigma} \} 
\Psi_{val} (x) \cdot
\Psi^{\dagger}_{val} (y) \vec{\tau} \Psi_{n} (y)}
{E_n - E_{val}} \right .
\nonumber \\  &+ & \left. 
\frac{1}{2} \sum_{m}
\Psi^{\dagger}_{n} (x)\gamma_{5} \{\hat{r} \times \vec{\sigma} \} 
\Psi_{m} (x) \cdot 
\Psi^{\dagger}_{m} (y) \vec{\tau} \Psi_{n} (y) 
{\cal R}_{\cal M} (E_n, E_m) \right ],
\nonumber \\
{\cal X}_2 (\vec{Q}^2) & = & N_c
\sum_{m^0} \int d^3 x j_1 (qr)
\int d^3 y \left[
\frac{\Psi^{\dagger}_{m^0} (x)\gamma_{5}
\{\hat{r} \times \vec{\sigma} \} \cdot \vec{\tau} 
\Psi_{val} (x) 
\Psi^{\dagger}_{val} (y) \Psi_{m^0} (y)}
{E_{m^0} - E_{val}} \right .
\nonumber \\  & & \hspace{1cm} 
\;+\; \left . \sum_{n}
\Psi^{\dagger}_{n} (x)\gamma_{5} \{\hat{r} \times \vec{\sigma} \} 
\cdot \vec{\tau} \Psi_{m^0} (x) 
\Psi^{\dagger}_{m^0} (y) \Psi_{n} (y) 
{\cal R}_{\cal M} (E_n, E_{m^0}) \right ],
\nonumber \\
{\cal M}_0 (\vec{Q}^2) & = &  \frac{N_c}{3} \sum_{n} 
\int d^3 x j_1 (qr)
\int d^3 y \left[ 
\frac{\Psi^{\dagger}_{n} (x) \gamma_{5}
\{\hat{r} \times \vec{\sigma} \} \cdot \vec{\tau}   
\Psi_{val} (x) 
\Psi^{\dagger}_{val} (y) \beta \Psi_{n} (y)}
{E_{n} - E_{val}} \right .
\nonumber \\  & & \hspace{1cm} 
\;+\; \left . \frac{1}{2} \sum_{m}
\Psi^{\dagger}_{n} (x)\gamma_{5} \{\hat{r} \times \vec{\sigma} \} 
\cdot \vec{\tau}  \Psi_{m} (x) 
\Psi^{\dagger}_{m} (y)\beta \Psi_{n} (y) 
{\cal R}_{\beta} (E_n, E_m) \right ],
\nonumber \\
{\cal M}_1 (\vec{Q}^2) & = & \frac{N_c}{3}
\sum_{n} \int d^3 x j_1 (qr)
\int d^3 y \nonumber \\  & \times &
\left[
\frac{\Psi^{\dagger}_{n} (x)\gamma_{5}
\{\hat{r} \times \vec{\sigma} \} 
\Psi_{val} (x) \cdot
\Psi^{\dagger}_{val} (y) \beta \vec{\tau} \Psi_{n} (y)}
{E_n - E_{val}} \right .
\nonumber \\  &+ & 
\left . \frac{1}{2} \sum_{m}
\Psi^{\dagger}_{n} (x)\gamma_{5} \{\hat{r} \times \vec{\sigma} \} 
\Psi_{m} (x) \cdot 
\Psi^{\dagger}_{m} (y) \beta \vec{\tau} \Psi_{n} (y) 
{\cal R}_{\beta} (E_n, E_m) \right ],
\nonumber \\
{\cal M}_2 (\vec{Q}^2) & = &  \frac{N_c}{3} \sum_{m^0} 
\int d^3 x j_1 (qr)
\int d^3 y \nonumber \\  & \times &
\left[ 
\frac{\Psi^{\dagger}_{m^0} (x) \gamma_{5}
\{\hat{r} \times \vec{\sigma} \} \cdot \vec{\tau}   
\Psi_{val} (x) 
\Psi^{\dagger}_{val} (y)\beta \Psi_{m^0} (y)}
{E_{m^0} - E_{val}} \right .
\nonumber \\  &+ & 
\left . \sum_{n}
\Psi^{\dagger}_{n} (x)\gamma_{5} \{\hat{r} \times \vec{\sigma} \} 
\cdot \vec{\tau}  \Psi_{m^0} (x) 
\Psi^{\dagger}_{m^0} (y) \beta \Psi_{n} (y) 
{\cal R}_{\beta} (E_n, E_{m^0}) \right ]  .
\label{Eq:mdens}
\end{eqnarray}
The regularization functions for the $G^{s}_M$ are 
\begin{eqnarray}
{\cal R} (E_n) & = & \int \frac{du}{\sqrt{\pi u}} 
\phi (u;\Lambda_i) |E_n| e^{-uE^{2}_{n}},
\nonumber \\
{\cal R}_{\cal Q} (E_n, E_m) & = & \frac{1}{2\pi} c_i
\int^{1}_{0} d\alpha \frac{\alpha (E_n + E_m) - E_m}
{\sqrt{\alpha ( 1 - \alpha)}} 
\frac{\exp{\left (-[\alpha E^{2}_n + (1-\alpha)E^{2}_m]/
\Lambda^{2}_i  \right)}}{\alpha E^{2}_n + (1-\alpha)E^{2}_m},
\nonumber \\
{\cal R}_{\beta}  (E_n, E_m) & = & 
\frac{1}{2\sqrt{\pi}} \int^{\infty}_{0} 
\frac{du}{\sqrt{u}} \phi (u;\Lambda_i) 
\left[ \frac{E_n e^{-uE^{2}_{n}} - E_m e^{-uE^{2}_{m}}}
{E_n - E_m}\right].
\label{Eq:regulm}
\end{eqnarray}
The cutoff parameter 
$\phi(u;\Lambda_i)=\sum_i c_i \theta 
\left(u - \frac{1}{\Lambda^{2}_{i}} \right)$ is  
fixed by reproducing the pion decay constants and other mesonic properties
\cite{Review}.
\end{appendix}
\begin{table}
\caption{The strange magnetic moments and mean-square strange radius
as varying the constituent quark mass.  
 The kaon tail shows a Yukawa mass of $\mu = m_{\rm K}\simeq 490$ MeV.
Our final values are in this table with $M=420$ MeV.}
\begin{tabular}{c|c|c|c|c|c|c}
$M$ & \multicolumn{2}{c|}{$400\; \mbox{MeV}\;\;\;\;\;\;\;\;$} 
& \multicolumn{2}{c|}{$420 \;\mbox{MeV}\;\;\;\;\;\;\;\;$}
&  \multicolumn{2}{c}{$450 \;\mbox{MeV}\;\;\;\;\;\;\;\;$}  \\ \cline{1-7}
$m_s$ [MeV] & $0$ & $180$ & $0$ & $180$ & $0$ & $180$ \\ \hline
$\mu_s[\mu_N]$ &$-0.66$ &$-0.69$& $-0.65$ & $-0.68$ & $-0.63$ &$-0.65$  \\
$\langle r^2 \rangle^{\rm Dirac}_{s} [\mbox{fm}^2]$ 
& $0.144$ & $-0.081$ & $0.120$ & $-0.051$ & $0.090$ & $-0.044$ \\
$\langle r^2 \rangle^{\rm Sachs}_{s} [\mbox{fm}^2]$ 
& $0.100$ & $-0.127$ & $0.077$ & $-0.095$ & $0.049$ & $-0.086$
\end{tabular}
\end{table}
\begin{table}
\caption{The strange magnetic moments and mean-square strange radius
as varying the constituent quark mass.  
The pion tail shows a Yukawa mass of $\mu=m_\pi=139$ MeV.}
\begin{tabular}{c|c|c|c|c|c|c}
$M$ & \multicolumn{2}{c|}{$400\; \mbox{MeV}\;\;\;\;\;\;\;\;$} 
& \multicolumn{2}{c|}{$420 \;\mbox{MeV}\;\;\;\;\;\;\;\;$}
&  \multicolumn{2}{c}{$450 \;\mbox{MeV}\;\;\;\;\;\;\;\;$}  \\ \cline{1-7}
$m_s$ [MeV] & $0$ & $180$ & $0$ & $180$ & $0$ & $180$ \\ \hline
$\mu_s[\mu_N]$ & $-0.81$ & $-0.42$ & $-0.78$ & $-0.44$ & $-0.74$ &$-0.50$  \\
$\langle r^2 \rangle^{\rm Dirac}_{s} [\mbox{fm}^2]$ 
& $-0.20$ & $-0.36$ & $-0.19$ & $-0.32$ & $-0.16$ & $-0.27$ \\
$\langle r^2 \rangle^{\rm Sachs}_{s} [\mbox{fm}^2]$ 
& $-0.25$ & $-0.39$ & $-0.25$ & $-0.35$ & $-0.21$ & $-0.31$
\end{tabular}
\end{table}
\begin{table}
\caption{The theoretical comparison for the strange magnetic moment 
and mean-square strange radius between different models.  
$M=420$ MeV, $m_s=180$ MeV, and $\mu=m_{\rm K}$ are 
used for the present work.}
\begin{tabular}{cccc} 
models&$\mu_s[\mu_N]$
&$\langle r^2\rangle^{\rm Sachs}_{s}[\mbox{fm}^2]$&references \\
\hline
Jaffe & $-0.31 \pm 0.09$ & $0.14\pm 0.07$ & \cite{jaffe} \\
Hammer {\em et al.} & $-0.24 \pm 0.03$ & $0.23\pm 0.03$ & 
\cite{Hammeretal} \\ 
Koepf {\em et al.} & $-2.6\times 10^{-2}$
&$-0.97\times10^{-2}$&\cite{Koepfetal} \\
Musolf $\&$ Burkhardt & $-(0.31\rightarrow 0.40)$
&$-(2.71\rightarrow 3.23)\times10^{-2}$& \cite{musolfa} \\
Cohen {\em et al.} &$-(0.24\rightarrow 0.32)$&$-(3.99\rightarrow 4.51)
\times 10^{-2}$ & \cite{Cohenetal} \\
Forkel {\em et al.} &\phantom{}&$1.69\times 10^{-2}$ &\cite{forkel} \\
Park $\&$ Weigel&$-0.05$&0.05&\cite{pw} \\
Park {\em et al.}& $-0.13$&$-0.11$&\cite{Parketal} \\
Leinweber & $-0.75\pm 0.30$ & &\cite{Leinweber} \\ 
Alberico {\em et al.} &$-0.14$&$0.055$ & \cite{Albericoetal} \\ 
Weigel {\em et al.}&$-0.05\rightarrow 0.25$ 
&$-0.2 \rightarrow -0.1$& \cite{tueb}\\
SU(3) $\chi$QSM &$-0.68$&$-0.095$  &Present work
\end{tabular}
\end{table}

\vfill\break
\begin{center}
{\Large {\bf Figure Captions}}
\end{center}
\noindent
{\bf Fig. 1}:
The strange electric form factor $G^{s}_{E}$ as functions of $Q^2$
with the $\mu = m_{\rm K}$:
The solid curve corresponds to
the constituent quark mass M=420 MeV, while
dot-dashed curve draws M=400 MeV.  The dashed curve
displays the case of M=450 MeV.  The M=420 MeV is distinguished 
since all other observables of the nucleon are then basically 
reproduced in this model. 
\vspace{0.8cm}

\noindent
{\bf Fig. 2}:
The strange electric form factor $G^{s}_{E}$ as functions of $Q^2$
with the $\mu = m_{\rm K}$:
 The solid curve corresponds to
the $m_s = 180$ MeV, while
dashed curve draws $m_s = 180$ MeV.  
The constituent quark mass $M$ is 420 MeV.
\vspace{0.8cm}

\noindent
{\bf Fig. 3}:
The strange electric form factor $G^{s}_{E}$ as functions of $Q^2$:
 The solid curve corresponds to the $\mu = m_{\rm K}$, while
dashed curve draws $\mu = m_\pi$.  
The constituent quark mass $M$ and $m_s$ are 420 MeV and 180 MeV,
respectively.
\vspace{0.8cm}

\noindent
{\bf Fig. 4}:
The strange electric density $r^2\rho^{s}_{E}$ as functions of $r$:
 The solid curve corresponds to the $\mu = m_{\rm K}$, while
dashed curve draws $\mu = m_\pi$.  
The constituent quark mass $M$ and $m_s$ are 420 MeV and 180 MeV,
respectively.
\vspace{0.8cm}

\noindent
{\bf Fig. 5}:
The strange magnetic factor $G^{s}_{M}$ as functions of $Q^2$
with the $\mu = m_{\rm K}$:
 The solid curve corresponds to
the constituent quark mass M=420 MeV, while
dot-dashed curve draws M=400 MeV.  The dashed curve
displays the case of M=450 MeV.  The M=420 MeV is distinguished 
since all other observables of the nucleon are then basically 
reproduced in this model. 
\vspace{0.8cm}

\noindent
{\bf Fig. 6}:
The strange magnetic form factor $G^{s}_{M}$ as functions of $Q^2$
with the $\mu = m_{\rm K}$: The solid curve corresponds to
the $m_s = 180$ MeV, while
dashed curve draws $m_s = 180$ MeV.  
The constituent quark mass $M$ is 420 MeV.
\vspace{0.8cm}

\noindent
{\bf Fig. 7}:
The strange magnetic form factor $G^{s}_{M}$ as functions of $Q^2$:
 The solid curve corresponds to the $\mu = m_{\rm K}$, while
dashed curve draws $\mu = m_\pi$.  
The constituent quark mass $M$ and $m_s$ are 420 MeV and 180 MeV,
respectively.
\vspace{0.8cm}

\noindent
{\bf Fig. 8}:
The strange magnetic density $r^2\rho^{s}_{M}$ as functions of $r$:
 The solid curve corresponds to the $\mu = m_{\rm K}$, while
dashed curve draws $\mu = m_\pi$.  
The constituent quark mass $M$ and $m_s$ are 420 MeV and 180 MeV,
respectively.
\vspace{0.8cm}

\vfill\break
\begin{center}
{\Large {\bf Figures}}
\end{center}
\vspace{1.6cm}
\centerline{\epsfysize=2.7in\epsffile{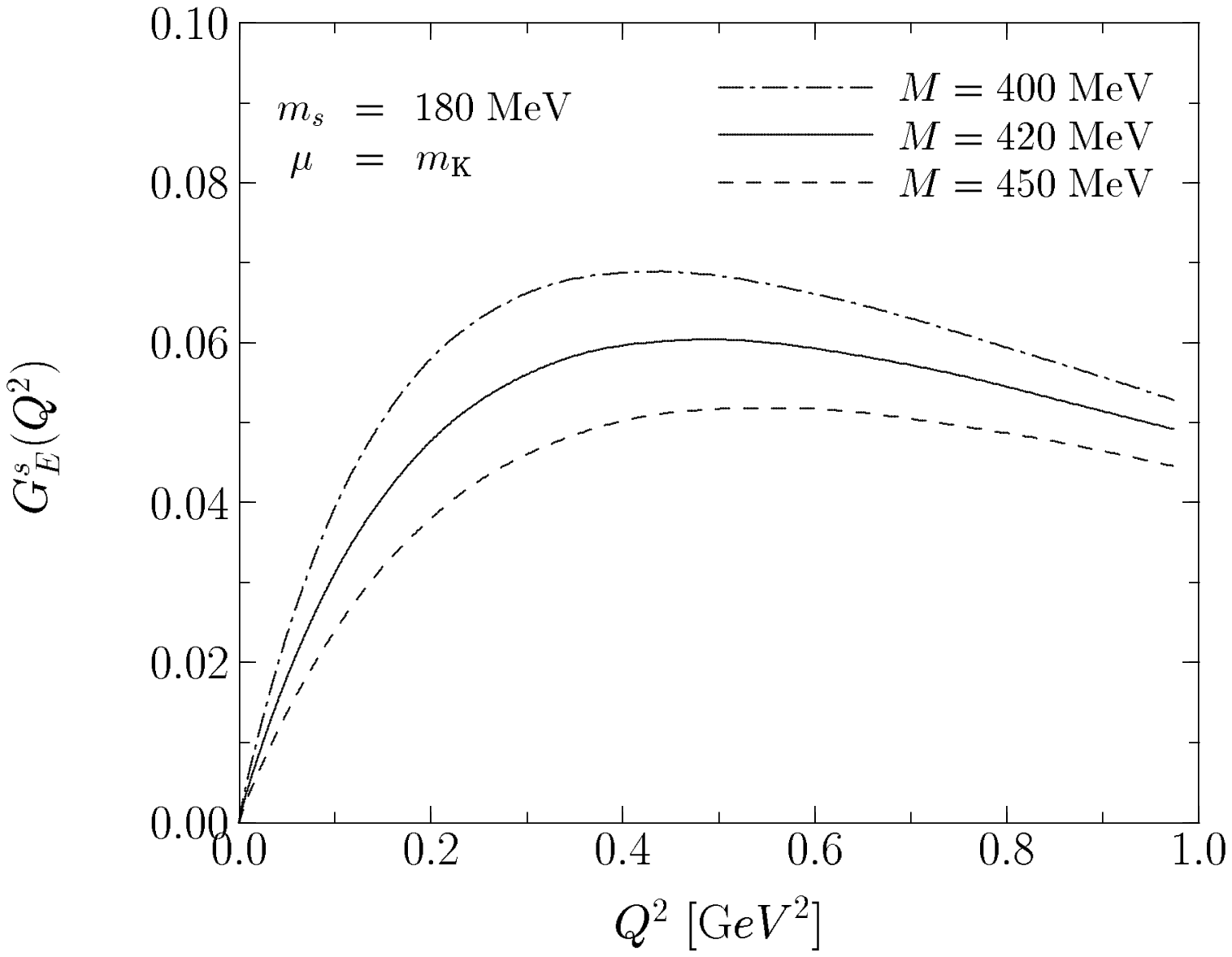}}\vskip4pt
\noindent \begin{center}	 {\bf Figure 1} 	 \end{center}   

\vspace{1.6cm}
\centerline{\epsfysize=2.7in\epsffile{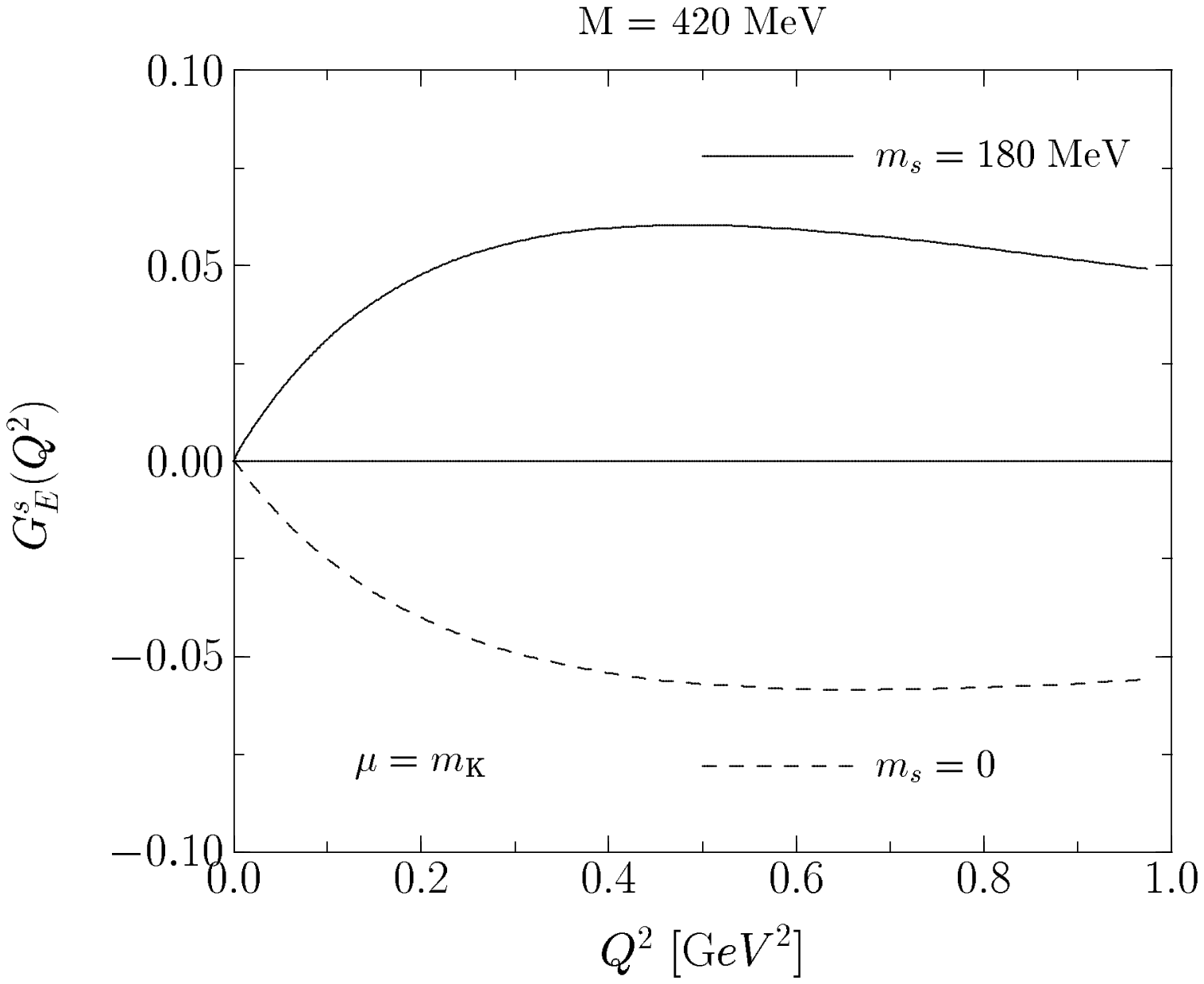}}\vskip4pt
\noindent \begin{center}	 {\bf Figure 2} 	 \end{center}   

\vspace{1.6cm}
\centerline{\epsfysize=2.7in\epsffile{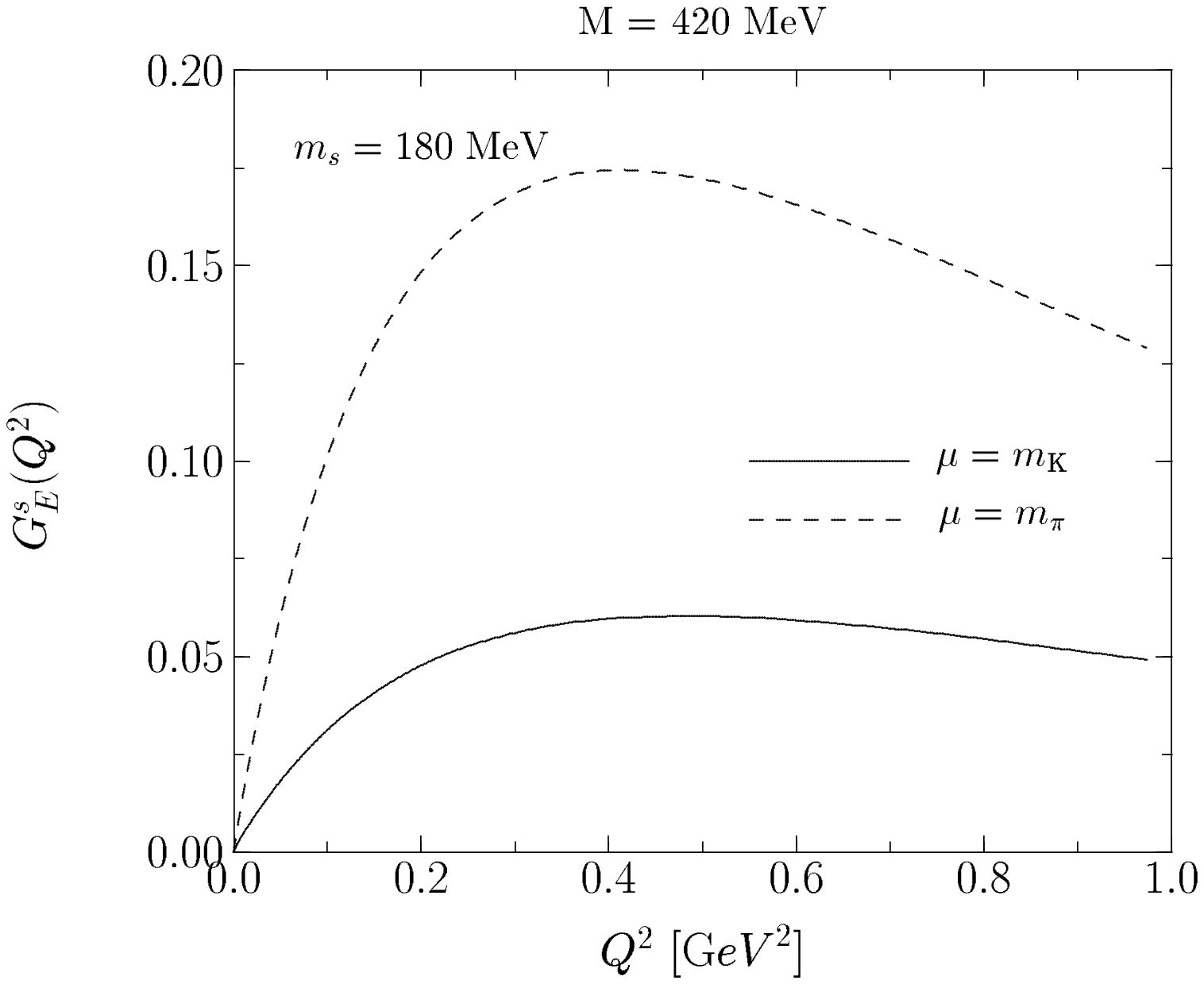}}\vskip4pt
\noindent   \begin{center}	 {\bf Figure 3} 	 \end{center}

\vspace{1.6cm}
\centerline{\epsfysize=2.7in\epsffile{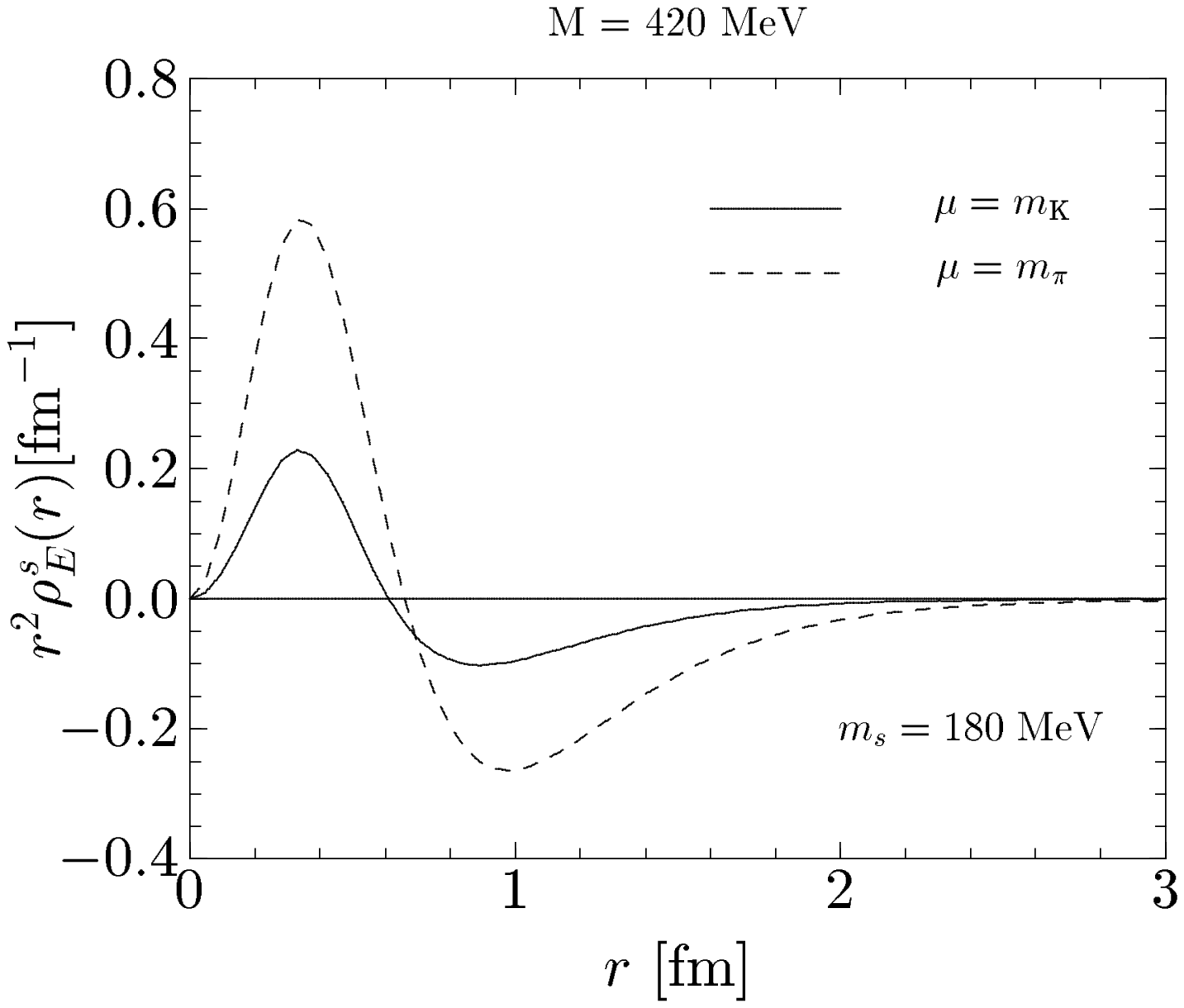}}\vskip4pt
\noindent   \begin{center}	 {\bf Figure 4} 	 \end{center} 

\vspace{1.6cm}
\centerline{\epsfysize=2.9in\epsffile{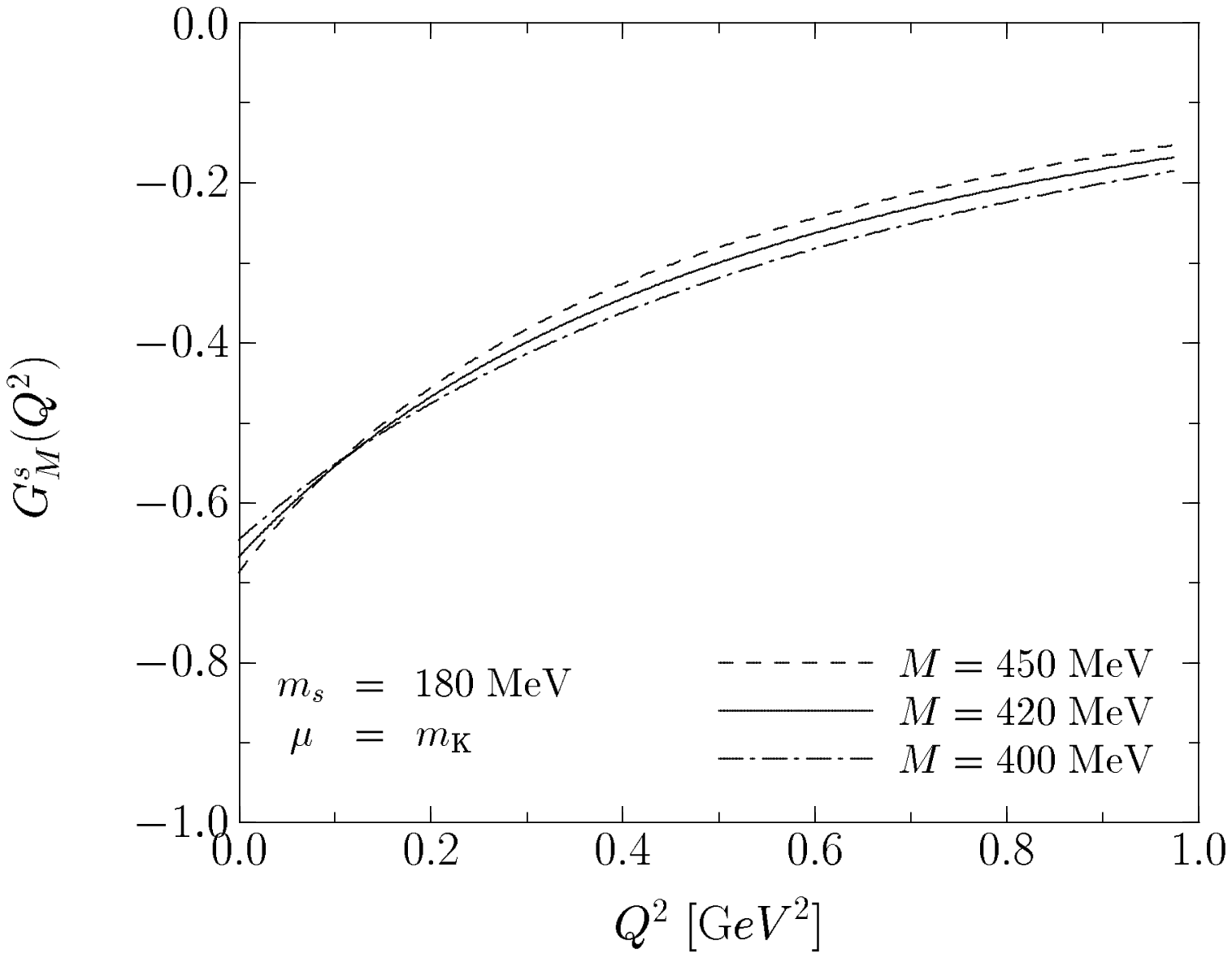}}\vskip4pt
\noindent   \begin{center}	 {\bf Figure 5} 	 \end{center} 

\vspace{1.6cm}
\centerline{\epsfysize=2.7in\epsffile{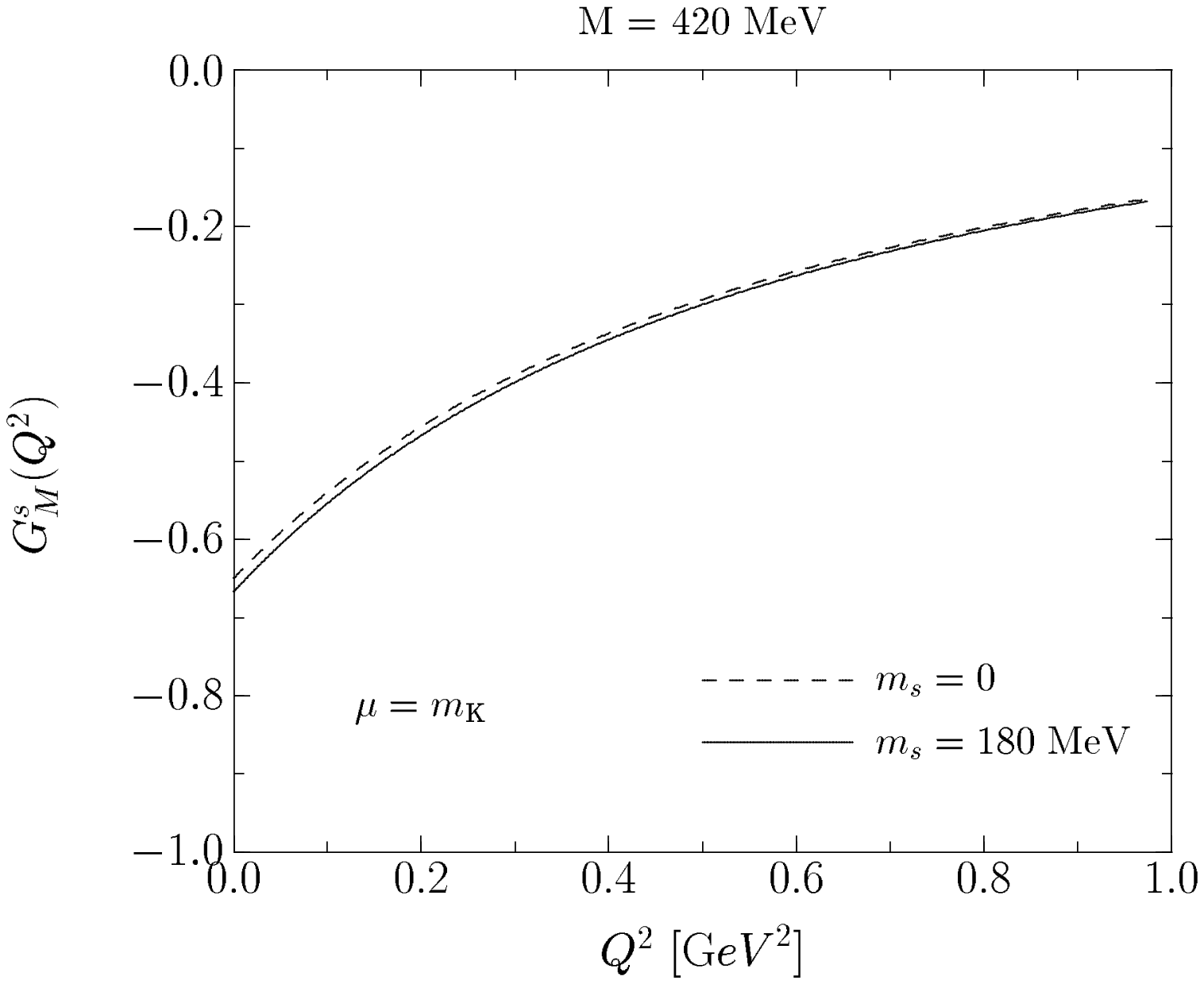}}\vskip4pt
\noindent   \begin{center}	 {\bf Figure 6} 	 \end{center} 

\vspace{1.6cm}
\centerline{\epsfysize=2.7in\epsffile{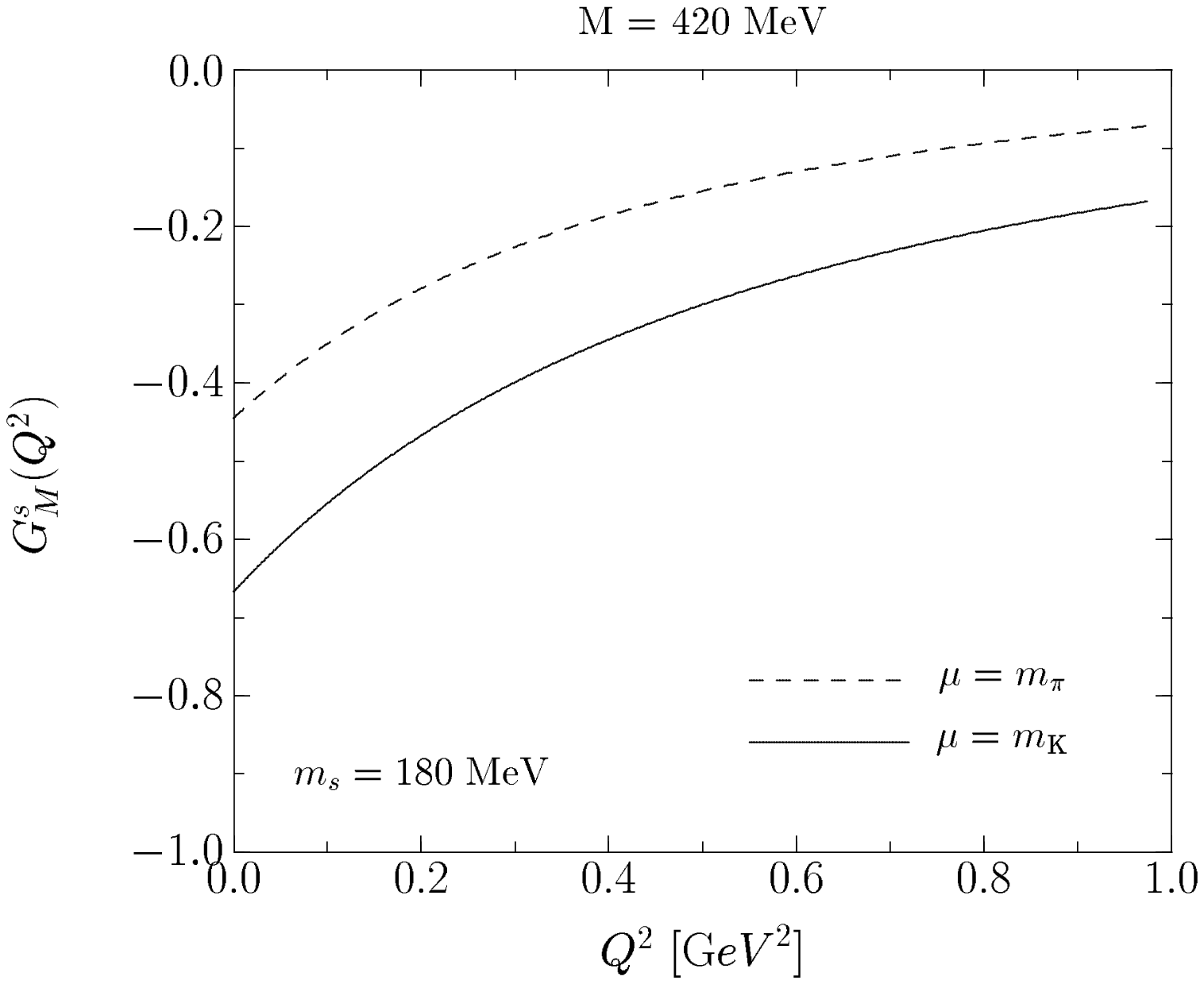}}\vskip4pt
\noindent   \begin{center}	 {\bf Figure 7} 	 \end{center} 

\vspace{1.6cm}
\centerline{\epsfysize=2.7in\epsffile{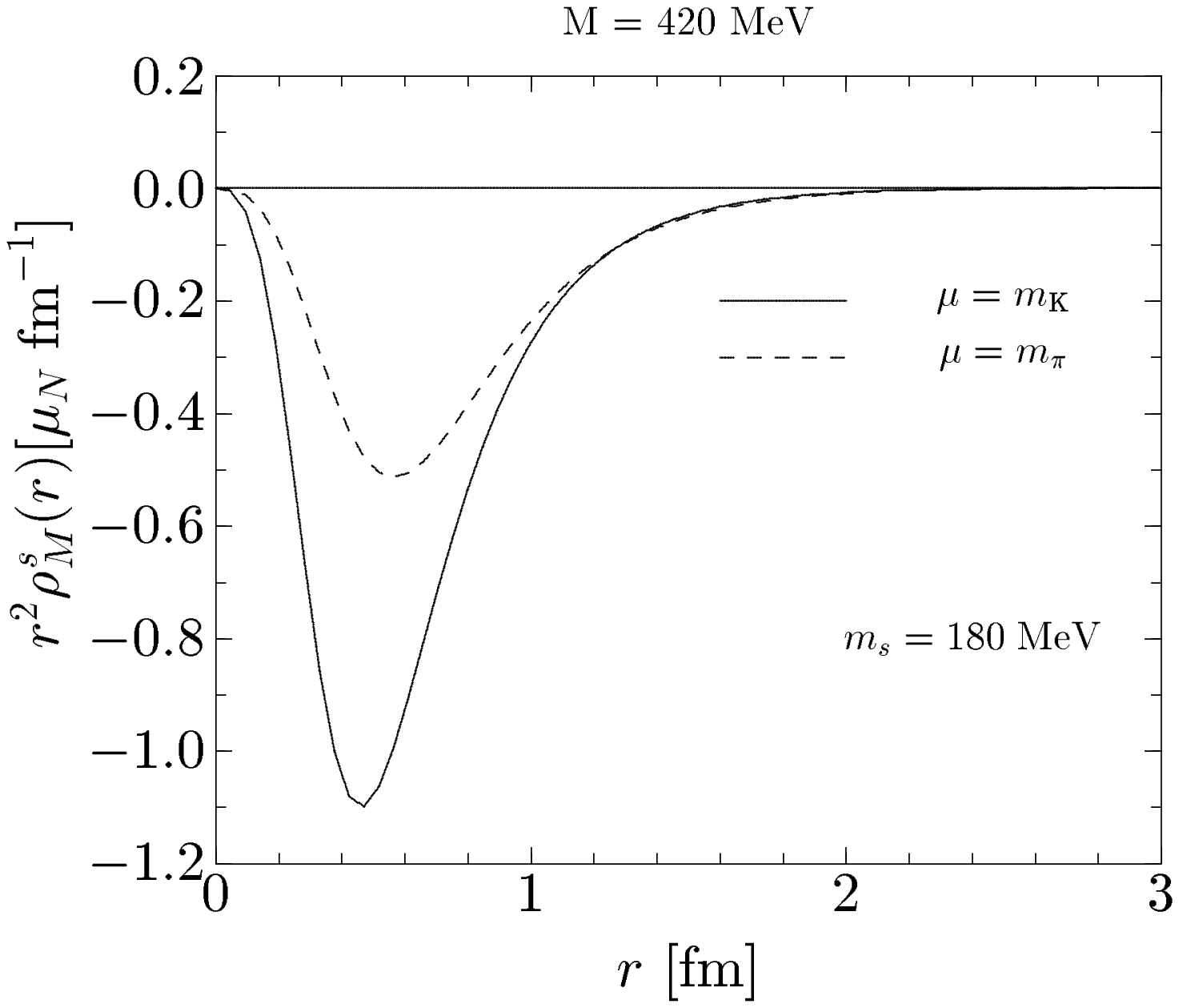}}\vskip4pt
\noindent   \begin{center}	 {\bf Figure 8} 	 \end{center}

\end{document}